\documentclass[aps,pre,twocolumn,superscriptaddress,floatfix]{revtex4}

\usepackage{amsmath, amsthm, amssymb}
\usepackage[normalem]{ulem}

\usepackage{latexsym}
\usepackage{amsfonts}
\usepackage{graphicx}
\usepackage{amsbsy}
\usepackage{float}
\usepackage{amsmath}

\begin{document}

\title{The cost of attack in competing networks}

\author{B.~Podobnik}
\affiliation{Center for Polymer Studies and Department of Physics,
  Boston University, Boston, MA 02215}

\affiliation{Faculty of Civil Engineering,
University of Rijeka, 51000 Rijeka,  Croatia}

\affiliation{Zagreb School of Economics and Management, 10000 Zagreb,
 Croatia}

\author{D. Horvatic}
\affiliation{Faculty of Natural Sciences, University of Zagreb,
 10000 Zagreb,  Croatia}

\author{T. Lipic}
\affiliation{Center for Polymer Studies and Department of Physics,
  Boston University, Boston, MA 02215}
\affiliation{Rudjer Boskovic Institute, Centre for Informatics and Computing, Zagreb, 10000, Croatia}

\author{M. Perc}
\affiliation{Faculty of Natural Sciences and Mathematics, University
of Maribor, Koro{\v s}ka cesta 160, 2000 Maribor, Slovenia}
\affiliation{Department of Physics, Faculty of Sciences, King
Abdulaziz University, Jeddah, Saudi Arabia}

\author{J. M. Buld\'{u}}
\affiliation{Center for Biomedical Technology (UPM),
 28223 Pozuelo de Alarc\'on, Madrid, Spain}

\affiliation{Complex Systems Group, Rey Juan Carlos University, 28933 M\'ostoles,
 Madrid, Spain}

\author{H.~E.~Stanley}
\affiliation{Center for Polymer Studies and Department of Physics,
  Boston University, Boston, MA 02215}

\begin{abstract}
Real-world attacks can be interpreted as the result of competitive interactions
between networks, ranging from predator-prey networks to networks of countries under
economic sanctions. Although the purpose of an attack is to damage a target network, it
also curtails the ability of the attacker, which must choose the duration and magnitude
of an attack to avoid negative impacts on its own functioning. Nevertheless, despite the
large number of studies on  interconnected networks, the consequences of initiating an
attack have never been studied. Here, we address this issue by introducing a model of
network competition where a resilient network is willing to partially weaken its own
resilience in order to more severely damage a less resilient competitor. The attacking
network can take over  the competitor nodes after their  long inactivity. However, due
to a feedback mechanism the takeovers weaken the resilience of the attacking network.
We define a conservation law that relates the feedback mechanism to the resilience
dynamics for two competing networks. Within this formalism, we determine the cost and
optimal duration of an attack, allowing a network to evaluate the risk of
initiating hostilities.
\end{abstract}

\maketitle

\section{Introduction}
Recent research carried out on competing interacting networks
\cite{Bascompte09,Bascompte14,Aguirre, dsouza,Arenas,Schweitzer} does not take
into account the fact that  real-world networks often compete not only to
survive but also to take over or even destroy their competitors
\cite{Thebault}.  For example, in international politics and economics,
when one country imposes economic sanctions on another, feedback
mechanisms can cause the country imposing the sanctions to also be
adversely affected.  The decision by a wealthier country to keep
military spending at a high level long enough to exhaust its poorer
competitor can also contribute to its own exhaustion
\cite{Richardson}. Similarly, in warfare, any attack depletes the
resources of the attacking force and can elicit a counter-attack from the
competing force \cite{Shakarian}.
Also, in nature, an incursion between
species can alter the dynamics of predator-prey interaction
\cite{Scheffer01}.

Although, these competing interactions are a widespread real-world
 phenomenon, current studies analyze only effects of attack on attacked
 networks, but disregarding its effect on the external attacking network. For
 example, for both single and interactive networks, existing studies on
 network robustness report that every network, regardless of the size and
 architecture, eventually can be destroyed
\cite{Podobnik15,Albert00,Cohen00,Reis_nphys14,buldyrev_n10,Dorogovtsev,PodobnikNSR2015}.
But, what then prevents a network from
  attacking a weaker competitor or,
  what is the optimal moment for initiating or ending an attack?
In order to identify
the factors that inhibit a  network from attacking and demolishing
a  weaker competitor and to determine the  optimal moment and duration  of an attack,
we develop a theoretical
framework that quantifies the cost of an attack by connecting the
feedback mechanisms and resilience dynamics between two competing
dynamic networks with differing levels of resilience
\cite{May2014,plosscheffer}.

\section{Theoretical framework}

\begin{figure*}[htb]
\centerline{\includegraphics[width=0.42\textwidth]{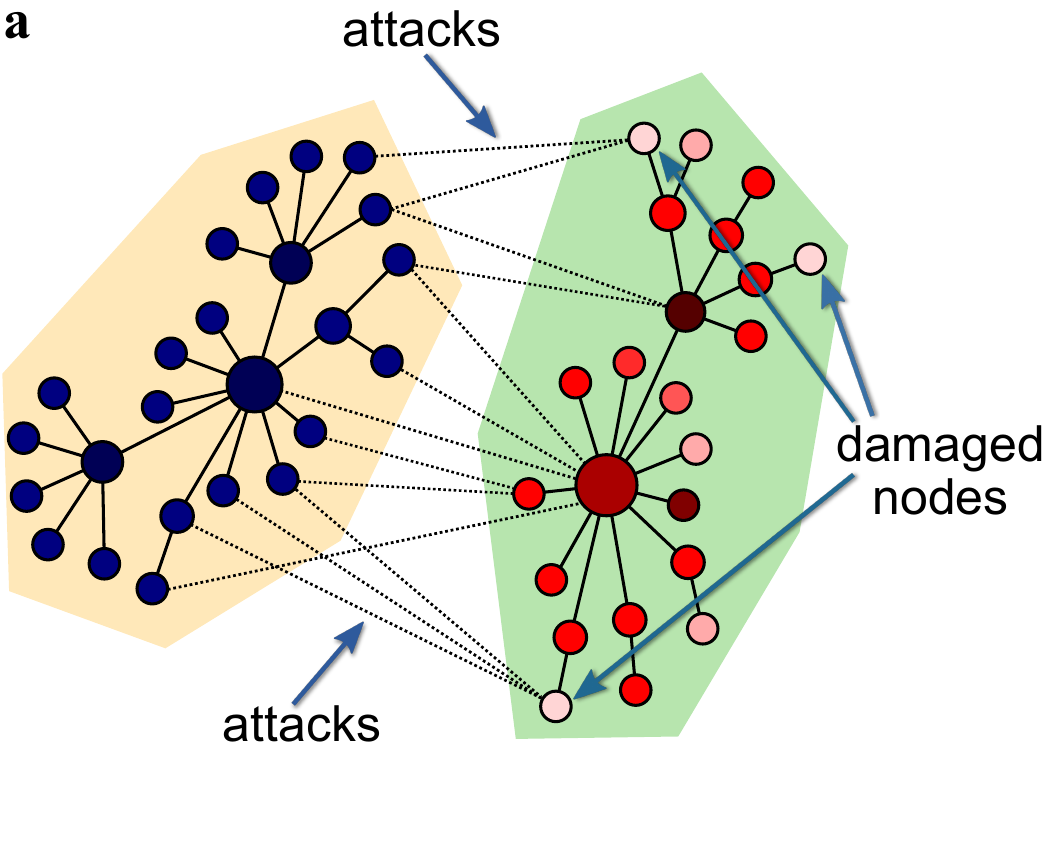}\hspace{8mm}
\includegraphics[width=0.42\textwidth]{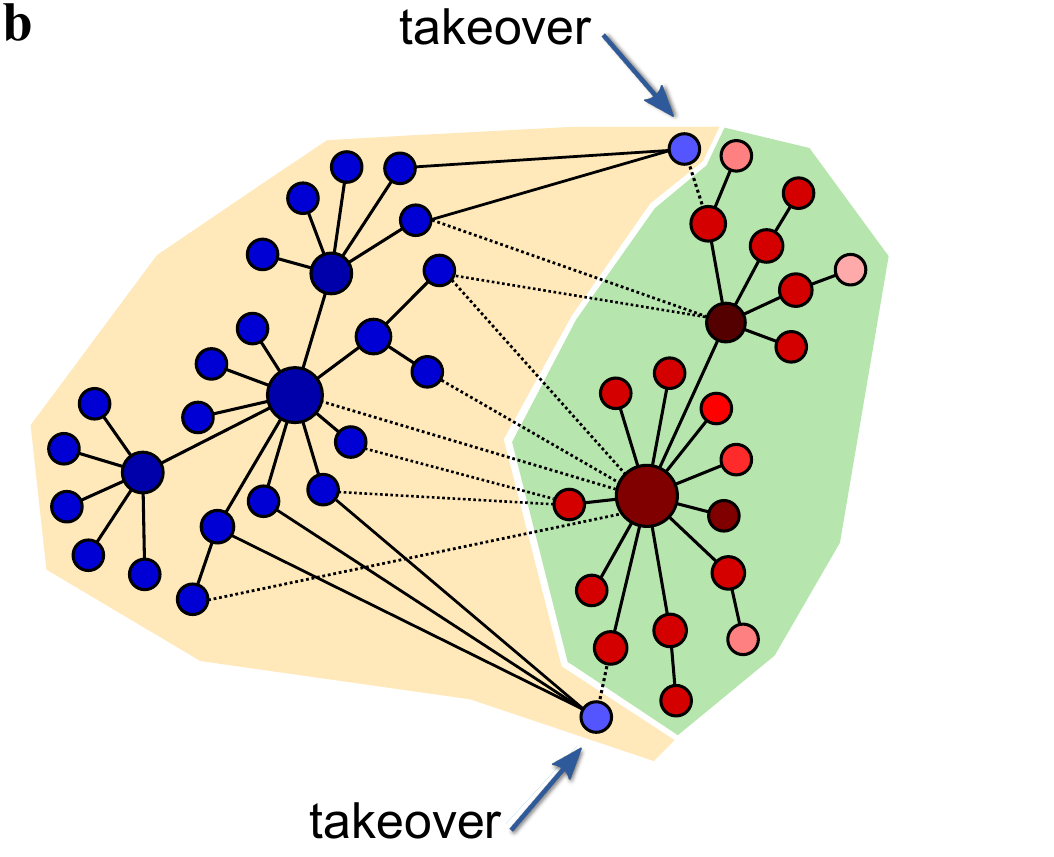}}
\caption{Attacks, failures, takeovers and their cost on the attacking network.
 In (a), we assume that
  each node in both the more resilient (stronger) network S and the less
  resilient (weaker) network W is described by the same failure
  probability.  Different nodes  spend different times during internal
  failure---the less opaque a node is, the more time it spends in
  internal failure.  In (b),  if a node in the weaker
  network W remains inactive more than some threshold time, it will be
  taken over by the stronger network S. However, network S pays for this
  takeover  with a reduction  of its resilience. }
\label{10}
\end{figure*}

We introduce a general methodology that can be applied to networks of
any size and structure.
 First, as an illustrative example,
 we describe two competing    Barab{\'a}si-Albert (BA) networks \cite{Barabasi99}
  which we
designate network S and network W.
 This model differs from the single network BA model in that the two interconnected networks
have both intra-network and inter-network links \cite{EPL12}. One
real-world example of this kind of network interaction is firms in an
economic network that link with other firms both domestically and
abroad.

Using the  preferential attachment
(PA) rule \cite{Barabasi99,EPL12,Perc14}, we generate networks S and W  starting with $n_0$ nodes
 in each network.
At each time step we
add a new node that connects with $m_S$ existing nodes in network S and
with $m_{W,S}$ existing nodes in network W, where the probability of
each connection depends on the total node degrees in networks S and W.
Similarly, using the PA rule we connect a new node in network W with
$m_W$ nodes inside network W and with $m_{S,W}= m_{W,S}$ nodes in network
S.

 In a broad class of real-world networks, nodes can fail either due to  inherent reason~\cite{Antonio04}
  or because their functionality depends on their neighborhood~\cite{Watts02,Antonio04}.
 Hence, any node in either of the two networks, e.g., a node $n_i$ inside network
S with $k_S$ neighbors in its own network and $k_{W,S}$ neighbors in
network W, can fail at any moment, either internally---independent of
other nodes---with a probability $p_{1}$ or externally with a
probability $p_{2}$.
Node $n_i$ externally fails with a probability $p_{2}$
 when, similar to the Watts model~\cite{Watts02},  the total fraction
of its active neighbors  is less than or equal to a fractional
threshold $T$ which is equal for all nodes in both networks.
 The larger
 the $T$ value, the less resilient the network. We assume that one
of the two networks is more resilient than the other, distinguishing
between strong network S and weak network W.  We do so
by assigning different fractional thresholds to the strong and weak
networks, $T_S$ and $T_W$, respectively, with $T_S < T_W$. As in
Ref.~\cite{Antonio04}, we assume that an internally-failed node in
network S or network W recovers from its last internal failure after a
period $\tau$.  Consecutive failures of the same node stretch the
effective failure times and introduce heterogeneity into the
distribution of inactivity periods. Since in real-world networks
 it is dangerous for nodes to be inactive,
 we allow the strong network
to take over nodes in the weak network when a node $n_i$ spends more time
in internal failure than $n \tau$, where $n$ is a
constant. Figure~\ref{10} qualitatively shows the interaction process.

\section{Results}

\begin{figure}[b]
\centering \includegraphics[width=0.23\textwidth]{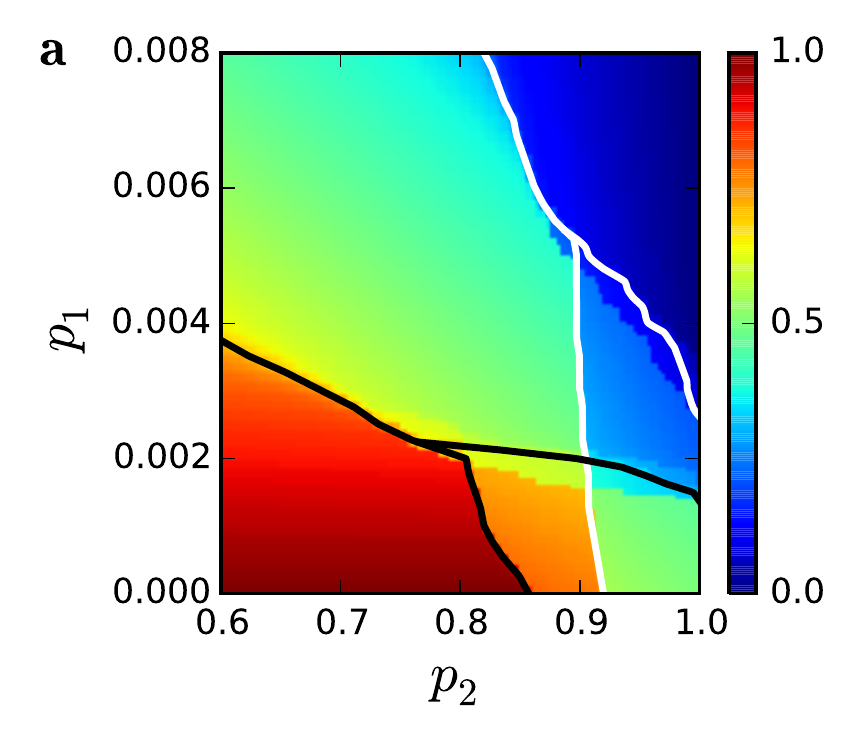}
\centering \includegraphics[width=0.23\textwidth]{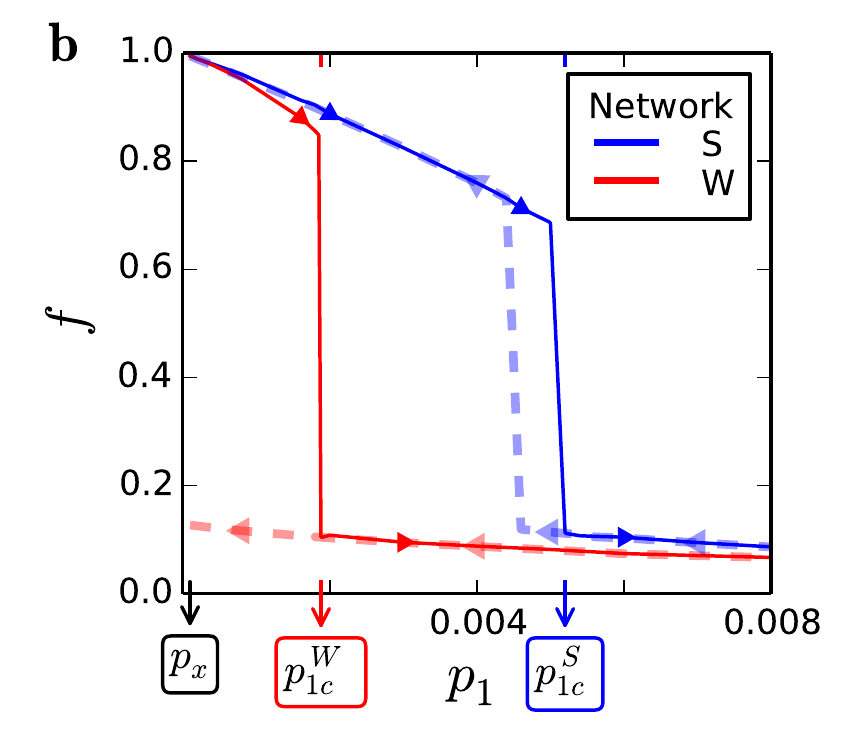}
\centering \includegraphics[width=0.23\textwidth]{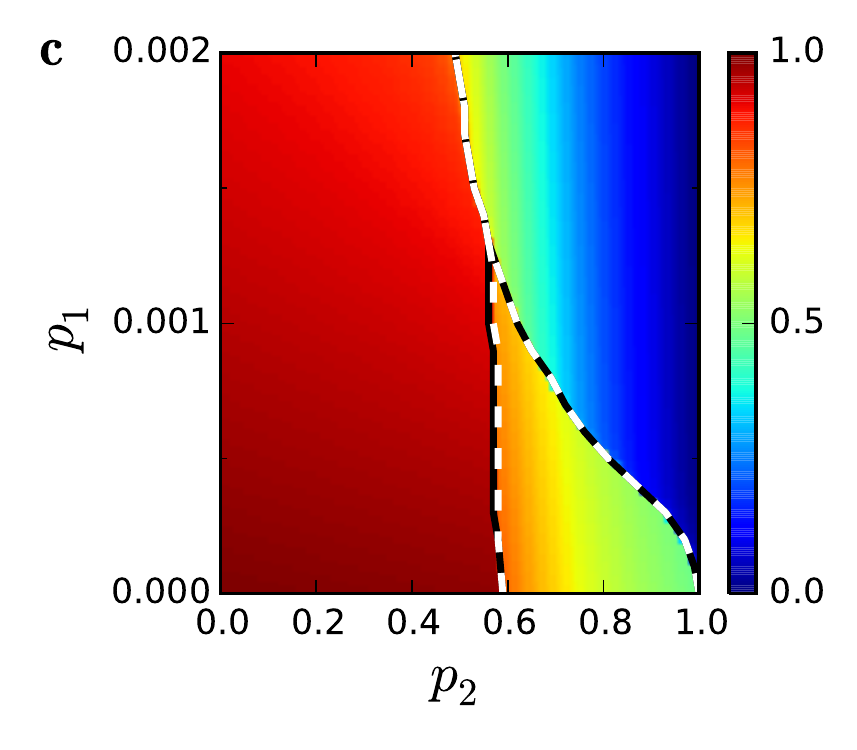}
\centering \includegraphics[width=0.23\textwidth]{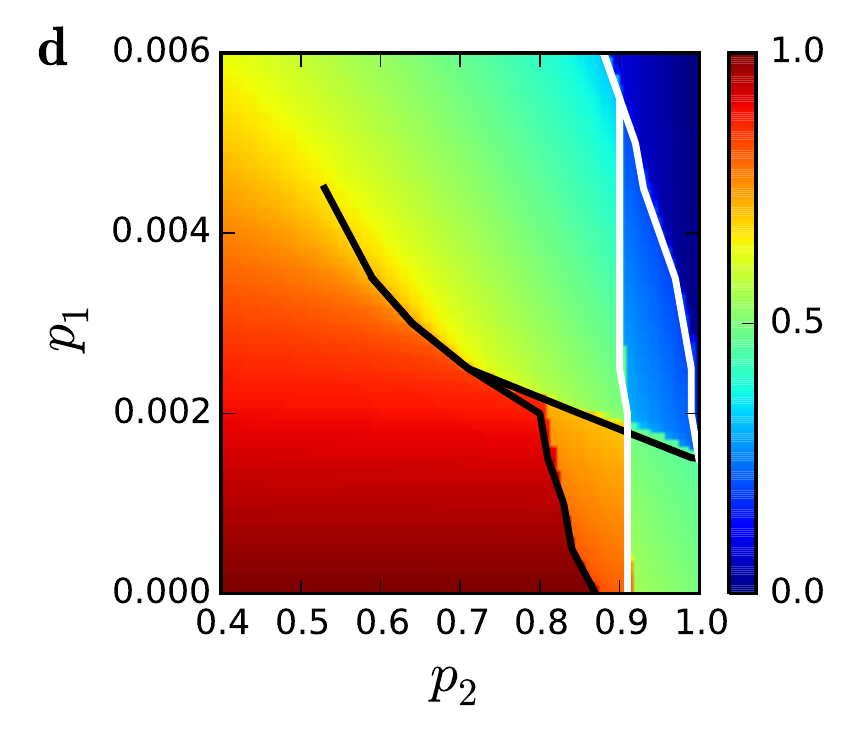}

\caption{ Attack strategy between two competing networks with
    different resilience levels and intra/inter link architecture.
  Shown are fractions of active nodes. The most resilient, strong
  network S (with $T_S=0.3$) endangers and partially destroys its own
  nodes by increasing their internal failure probability $p_1$ in order
  to more severely damage the least resilient network W (with
  $T_W=0.7$). Each of S and W has the hysteresis composed of two
  spinodals, representing attacking and recovery phase.  The recovery
  time is $\tau = 50$ and the takeover and cost mechanisms are
  disregarded.  (a) Attacking strategy between two competing BA
  networks with parameters: $m_S=m_W= 3$ and $m_{S,W}=m_{W,S}=2$. Strong
  network S wants to bring W in parameter space between hystereses of W
  (black lines) and S (white lines) where S is predominantly active and
  W is predominantly inactive (see, b). Dark red (blue) is parameter
  space where both S and W are active (inactive).  (b) For
  $p_2=0.9$, fraction of active nodes in the strong $f_S$ (blue lines)
  and weak $f_W$ (red lines) networks as a function of the internal
  failure probability $p_1$.  Hysteresis is a result of increasing $p_1$
  from zero to one and then decreasing it back to zero. The increase of
  $p_1$ accounts for the attacks and the decrease for a repair of the
  network.  (c) Same case as (a) but for two randomly connected
  competing Erd\H{o}s-Renyi networks.  (d) Same case as (c) but
  with an assortative mixing in the connection between networks: nodes
  with degree $d_1$ link, with probability $1/|d_1 - d_2 +1|$, with
  nodes in the other network with degree $d_2$.  }
\label{1}
\end{figure}

We quantify the current collective state of the strong and weak networks
in terms of the fraction of active nodes, $f_{S}$ and $f_{W}$,
respectively \cite{Antonio04,Pod14,Pod14_2}.  We assume that initially on both
networks have internal and external failure probability values of $p_{1}
\equiv p_X$ and $p_2$, respectively.  Figure~\ref{1}(a) shows a
two-parameter phase diagram for each network in which the hysteresis is
composed of two spinodals separating two collective states, i.e., the
primarily ``active'' and the primarily ``inactive.''  Figure~\ref{1}(b)
shows that increasing the value of $p_1$ leads to catastrophic
first-order phase transitions in both networks.  When each network
recovers (i.e., when $p_1$ is decreased to previous values), the
fraction of active nodes returns to an upper state. Nevertheless, the
critical point in the recovery is well beyond the point at which the
network collapses. Figure~\ref{1}(b) also shows (solid line) that the
initial choice of parameters makes network S more resilient to network
fluctuations in the value of $p_1$ and that the fluctuation needed to
initiate the collapse of network S ($p_1^S \equiv p_{1c}^S - p_X$) is
much larger  than the fluctuation needed to initiate the collapse of
network W ($p_1^W \equiv p_{1c}^W - p_X$).  Furthermore, network W is
closer to a critical transition than network S.

Because network S has a higher resilience than network W and can more
easily withstand fluctuations, S could induce the collapse of W by
increasing $p_1$, but only if the fraction of its active links is not
dramatically reduced.  Figure~\ref{1}(b) shows how when network S
attacks network W by increasing $p_1$ to $\approx 0.002$ the weak
network becomes abruptly dysfunctional. Figure~\ref{1}(b) also shows
that when the values of $p_1$ are reset to their pre-attack levels the
collapse of network W is permanent (red dashed line) and, if it ceases
its attack, the recovery of network S is complete and all of its
inactive nodes are reactivated (see blue dashed line  Figure 2(b)).
Similarly, when economic sanctions in a financial system are lifted the
weak economies are not restored but the strong economics recover after
suffering little damage.

Figure~\ref{1}(c) shows a modified competing network structure in which
there are two interconnected Erd\H{o}s-Reny networks \cite{ER}
  with inter-network
links randomly chosen. Although this structure quantitatively differs
from the phase diagram of competing BA networks, the same kind of
transition occurs in the random configuration. This indicates the
generality of these critical transitions in competing networks. We
obtain similar results when degree-degree correlations are introduced
between the links connecting both networks. Figure~\ref{1}(d) shows
nodes in the strong network linking with nodes in the weak network only
when they are of similar degree (i.e., ``assortative mixing''
\cite{NW}). As in the other configurations, the better position of the
attacker enables the strong network to destroy the weak one and then
return safely to its initial state.

\subsection{Mean field theory}

Using mean-field theory we analytically describe the attack-and-recovery
process between two interconnected networks with random regular
topologies where all nodes within the same network have the same degree.
 We assume that each node in network S is linked with $k_S$
nodes in its own network and $k_{W,S}$ nodes in network W.  Similarly,
each node in network W is linked with $k_W$ nodes inside network W and
$k_{S,W}$ nodes in network S. In both networks the fraction of failed
nodes is $a \equiv 1 - f$, where $f$ is the fraction of functional
nodes.  We can approximate the values of $a$ at each network by
\begin{eqnarray}
 a_S &=& p^*_{S,1} + p_{S,2} (1 -p^*_{S,1}) E_S  \\
 a_W &=& p^*_{W,1} + p_{W,2} (1 -p^*_{W,1}) E_W,
 \label{a}
\end{eqnarray}
where $p^*_{S,1} \equiv 1 - \exp(- p_{S,1} \tau)$~\cite{Antonio04} denotes the average
fraction of internally failed nodes and
$p_{S,2}E_S$ denotes the probability that a node in network S has
externally failed,
\begin{eqnarray}
\nonumber
  E_S &=& \sum_{j=0}^{t_S}  \sum_{i=0}^j  {k_S \choose k_S -  i}
  a_S^{k_S -i} (1 - a_S)^i   \\
 && {k_{W,S} \choose k_{W,S}  - (j - i)}
  a_W^{k_{W,S} -(j -i)}  (1 - a_W)^{j-i}.
 \label{a2}
\end{eqnarray}
Here $t_S$ represents the absolute threshold of network S simply related to
the fractional threshold $T_S$  as $T_S=t_S/(k_S + k_{W,S})$: a node in
network S can externally fail with a probability $p_{S,2}$ only when the
number of active neighbors in both network S and network W is  less than or equal to
$t_S$. Similarly, we obtain $E_W$ for network W by
 replacing S with W, and vice versa, in Eq.~(\ref{a2}). Finally, we set network S to be more
 resilient than network W, by setting
$t_S/(k_S + k_{W,S}) < t_W/ (k_W + k_{S,W})$.

\begin{figure}[b]
\centering \includegraphics[width=0.23\textwidth]{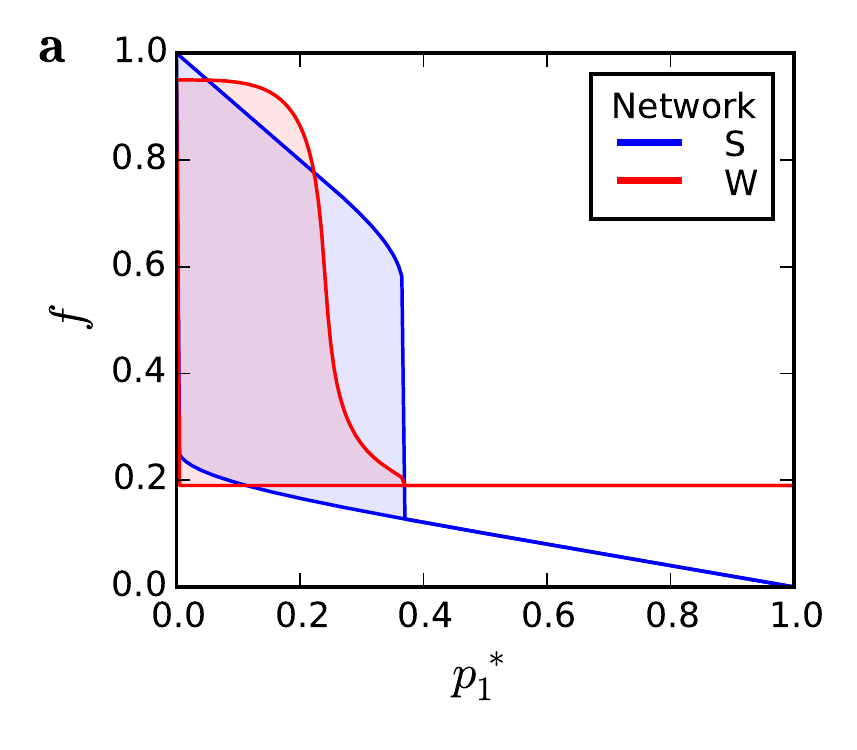}
\centering \includegraphics[width=0.23\textwidth]{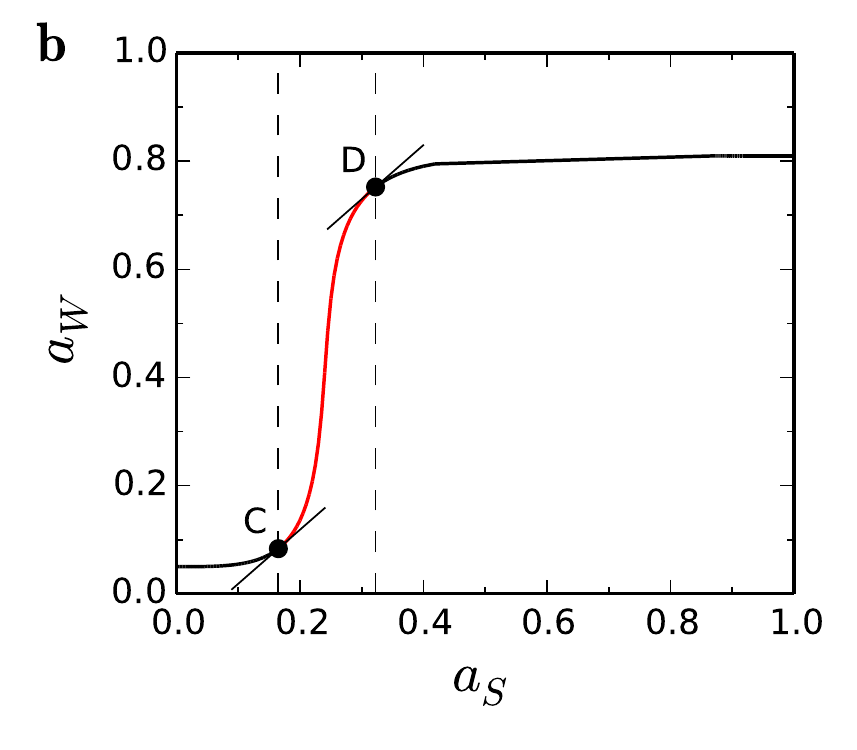}
\caption{ Identifying the optimal parameters for attacking a weak
    network.  Analytical approach. Strong network S (less vulnerable)
  uses its own probability of internal failures $p^*_{S,1}$ to cause
  damage in weak network W and, unavoidably, induce a partial
  self-destruction.  Model parameters are: $k_S = 20$, $k_{S,W} =
  k_{W,S} = 10$, $k_S = 5$, $t_S =10$, $t_W = 10$, $p_{S,2} = p_{W,2} =
  0.8$, and $p^*_{W,1} = 0.05$.  In (a), fraction of active nodes in
  network S and W, $f_{S}=1-a_S$ and $f_{W}=1-a_W$, respectively.
  Strong network S (blue) deliberately initiates its own failures
  (increasing $p^*_{S,1}$) to create larger damage in a weak (more
  vulnerable) network W (red).  Note that the fraction of active nodes
  exhibits a hysteresis behavior for both networks, with a critical
  point at $p_C \approx 0.33$.  In (b), we investigate when S should stop
  attacking W by increasing its probability of internal failure
  $p^*_{S,1}$. Shown are the fractions of failed nodes, $a_{S}=1-f_S$ and $a_{W}=1-f_W$.
   Between points C and D (dashed lines), an increase of
  $p^*_{S,1}$ induces more failures in the weaker network, leading to a
  comparative benefit. Beyond point D, the attack is not worthwhile for
  network S since it suffers the consequences more intensely than its
  competitor. }
\label{2}
\end{figure}

\begin{figure}[b]
\centering \includegraphics[width=0.23\textwidth]{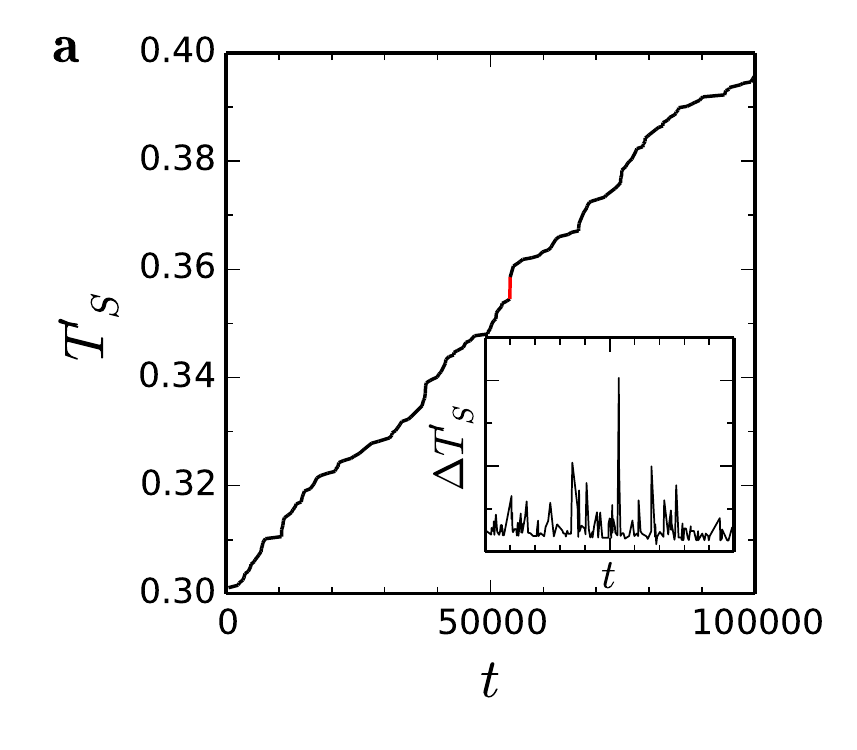}
\centering \includegraphics[width=0.23\textwidth]{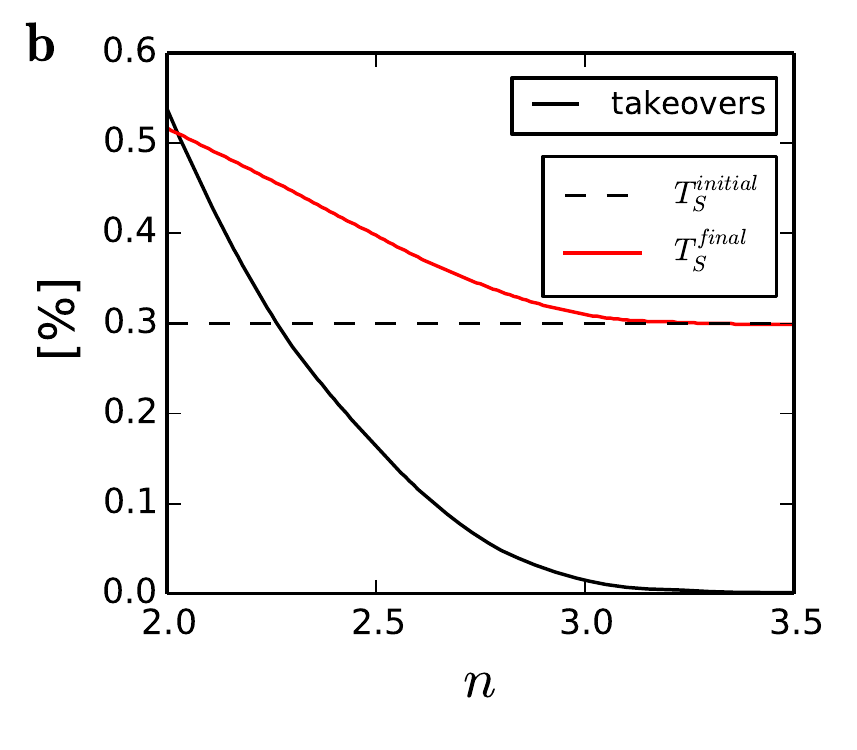}
\centering \includegraphics[width=0.23\textwidth]{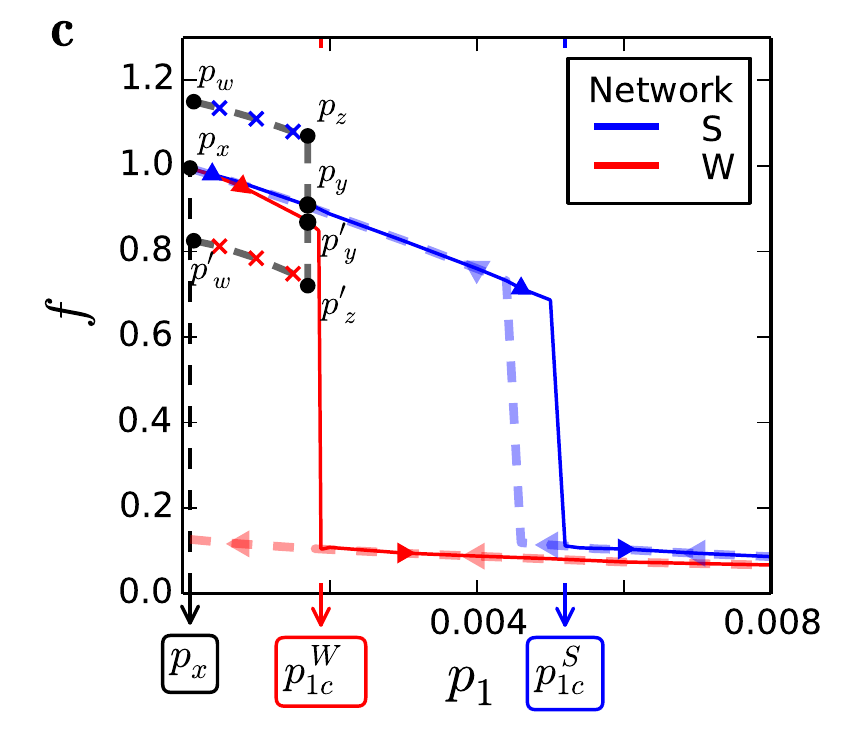}
\caption{  Cost and takeover mechanisms in two competing $BA$
    networks.  (a) Threshold $T'_S$ of network S as a function of
  time for two competing $BA$ networks with $n=2.5$ and $\tau= 50$.
  Fluctuations in the evolution of $T'_S$ are a consequence of the
  degree of the acquired node: the higher the degree the higher the
  increase of $T'_S$.  (b) Fraction of takeovers and final
  threshold as a function of the time $n \tau$ required to acquire a
  node from the weak network, with $\tau=50$.  As $n$ increases, the
  number of takeovers decreases to zero.  At the same time, the
  resilience of network S tends to the initial value $T_S=0.3$.  (c) Due to takeovers the fraction of active nodes in the more resilient network S can increase to values higher than one. In this example, network W is irreversibly damaged after $p_1$ is restored to its initial value.  }
\label{33}
\end{figure}

The analytical results of Fig.~\ref{2}(a) indicate that when network S
increases the internal failure probability $p_{S,1}$ and so $p^*_{S,1}$
 in an effort to
damage network W it also causes partial damage to itself. Although it first
seems that increasing $p^*_{S,1}$ reduces more active nodes in network S
than in network W, when $p^*_{S,1} > 0.18$ the fraction of active nodes in
network W drops sharply and eventually $f_S>f_W$. This attack strategy
by network S is thus effective. If $p^*_{S,1} > 0.33$, however, network S
undergoes a first order transition that leads to collapse, a situation
that network S clearly must avoid.

Inspecting the recovery of the previous internal failure probability
values after the attack we find that the fraction of active nodes in
both networks exhibit a hysteresis behavior. Note that when the
transition at $p^*_{S,1}\sim 0.33$ is surpassed neither network is able
to restore its functioning to the levels previous to the attack.

The analytical results indicate that attacking network S is effective
only for certain values of $p^*_{S,1}$.  Thus network S should increase $p^*_{S,1}$
only as long as the damage to network W continues to be greater than the
damage to itself, i.e., only when $\Delta a_W > \Delta a_S$.
Figure~\ref{2}(b) shows the region in which attacks by network S are
effective by showing the fraction of failed nodes in both networks in a
two-dimensional phase space as the value of $p^*_{S,1}$ is increased.  Two
solid lines with a slope of one indicate the region in which an attack
by network S is effective.  When the slope of function $a_W=f(a_S)$ is
greater than one (the region between the two shaded lines), increasing
$p^*_{S,1}$ produces more damage in network W than in network S and is thus an
effective attack strategy.

In order to measure the effect of capturing nodes from a competitor
network and how takeovers can modify the resilience properties of a
network, we design a model in which network S is again more resilient than
network W ($T_S < T_W$) and where node $n_i$ of network W is taken over by
network S if its internal failure time exceeds $n \tau$, where $\tau$ is
a certain failure time and $n$ a constant. Note that the longer a node
in network W remains inactive (i.e., the higher the value of $n$), the
higher the probably that it will be acquired by network S.
Real-world
examples of this mechanism include sick or disabled prey in an
ecological system \cite{Errington,PLOS}  or countries whose economic systems remain in
recession for too long a period.

\subsection{Take over and conservation laws}

To evaluate the acquisition costs in both networks we define network
wealth (capital) as proportional to two variables: the total number of
links in the network---as defined in conservation biology
\cite{Costanza,Hunter}---and the resilience of the network.  Note that
if two networks have the same number of links but different resiliencies
their wealth is not equal. Note also that when network S acquires a node
of degree $k_{W,i}$ from network W the overall resilience of network S
decreases because it has acquired a weaker node. Thus network S pays a
instantaneous, collective cost through a feedback mechanism that
decreases its resilience from an initial threshold $T_S$ to a new
threshold $T'_S$.

One of the important issues in dynamic systems that have a critical
point as an attractor is whether a conservation of energy is required in
local dynamic interactions \cite{Bak,dsouza13,Markovic}.  To quantify
how threshold $T'_s$ changes in competing networks, we define a
conservation law that relates the feedback mechanism to the resilience
dynamics as
\begin{equation}
          N~\langle k_S \rangle ~(T'_S - T_S) = k_{W,i}(T_W - T'_S).
\label{cost}
\end{equation}
Here $N$ is the size of the strong network, $\langle k_S \rangle$ its
average degree, and $k_{W,i}$ the degree of the node that has been taken
over. Thus, we assume that the more important the acquired node (i.e.,
the larger its degree $k_{W,i}$), the greater the cost to the resilience
of network S, making it more vulnerable to future attacks.  As a result, when a
predator (strong) network S increases its size $N$ and its degree $\langle k_s
\rangle$, its acquisition cost, $T'_S - T_S$, will decrease.

Here we quantify how  threshold $T'_S$ of the stronger network
   changes in competing networks where
 we assume that threshold $T_W$ of the weaker network does not change
   because every node has the same threshold. The stronger network S has
    the initial number of nodes $N_S$,  the average degree $\langle k_S \rangle$.
 After a multiple takeovers, where $S$ took over nodes
  $n_{w,1}$, $n_{w,2}$, ..., $n_{w,n}$ with degrees
    $k_{w,1}$, $k_{w,2}$, ..., $k_{w,n}$, respectively,  by using
     Eq.~(4)  we
     obtain
     \begin{equation}
    T'_S  = \frac{(k_{w,1} + k_{w,1} +...+ k_{w,n})T_W + N_S \langle k_S \rangle T_S}{N_S \langle k_S \rangle + k_{w,1} + k_{w,1} +...+ k_{w,n}}.
\end{equation}


Figure~\ref{33}(a) shows that when network S acquires nodes in network W
the threshold $T'_S$ of network S is increasingly affected as time
passes.  In this example, a node in network W is taken over by
network S when the node is in failure state longer than $n \tau$ time
steps, where $n=2.5$ and $\tau=50$.  Note that as network S acquires
weak nodes, its threshold $T'_S$ increases and it becomes more
vulnerable.  Figure~\ref{33}(b) shows the interplay between the time
required to acquire a node $n \tau$ and the threshold $T'_S$.  Note that
as $n \tau$ increases, takeovers become increasingly rare and the final
threshold of network S approaches its initial resilience, here
$T_S=0.3$.

Figure~\ref{33}(c) shows that, if the example in Figure~\ref{1}(b) is
extended to include a takeover mechanism, a fraction of active nodes
$f_S$ in network S---measured relative to the initial number of nodes in
each network---reaches values higher than one, with a peak at $p_y
\rightarrow p_z$.  Note that when attacks cease (e.g., when, in an
economic system, sanctions are lifted) decreasing the value of $p_1$,
$p_z \rightarrow p_w$, the fraction of active nodes in network S
increases but network W is left irreversibly damaged (see the closed
hysteresis $p'_y \rightarrow p'_z \rightarrow p'_w$).

\subsection{Threshold diversity in competing networks}

Thus far we have studied competing interconnected networks in which there is only one
threshold characterizing each network. However, in real-world interconnected
 networks
 commonly the functionality of a node in a given network is not equally sensitive
  on  the neighbors in its own and the other network.
 To this end, we assume that node $n_i$ in network S can
externally fail with probability $p_2$ if the fraction of the active
neighbors of node $n_i$ in network S is equal to or lower than some
threshold $T_S$, or if the fraction of the active neighbors of node $n_i$
in network W is equal to or lower than some threshold $T_{W,S}$.  We
similarly define external failure in the less resilient network W by
replacing threshold $T_S$ with $T_W$. The functioning of each node is
thus dependent on its neighbors in network S and network W, but with
different sensitivities---different resilience to external fluctuations

\begin{figure}[b]
\centering \includegraphics[width=0.235\textwidth]{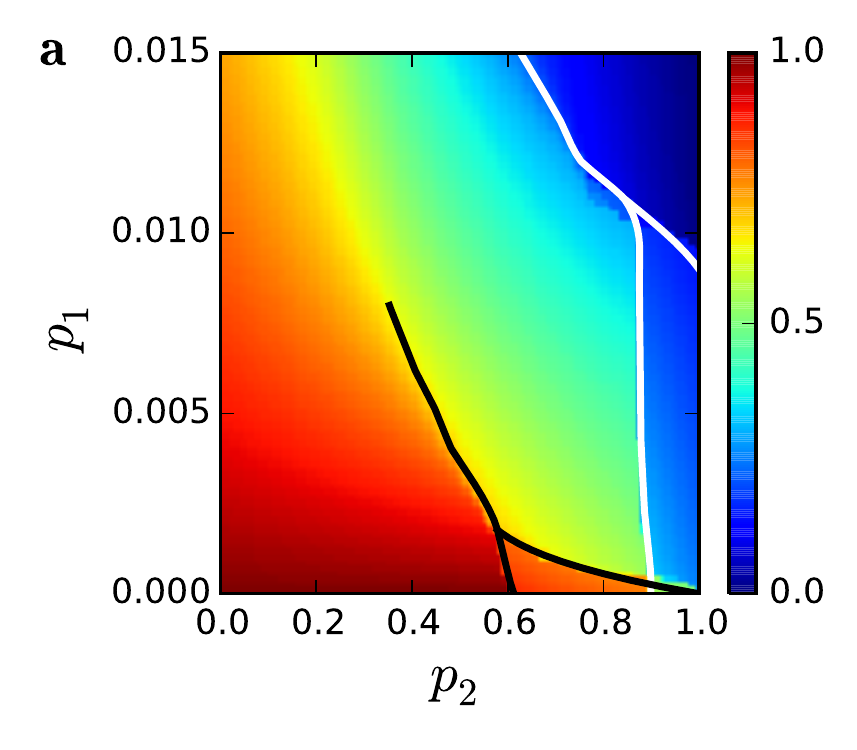}
\centering \includegraphics[width=0.23\textwidth]{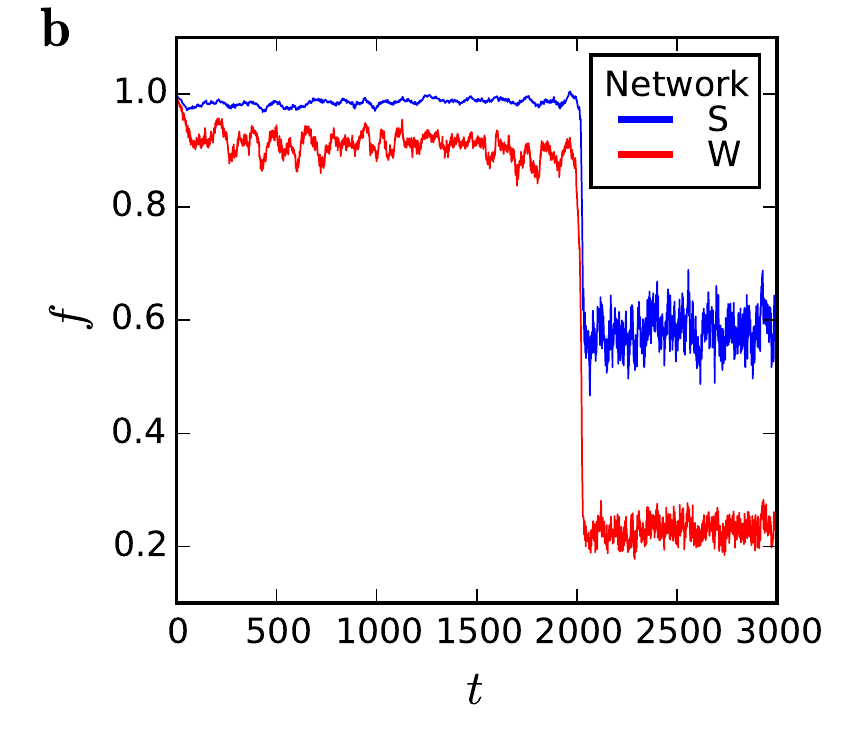}
\centering \includegraphics[width=0.23\textwidth]{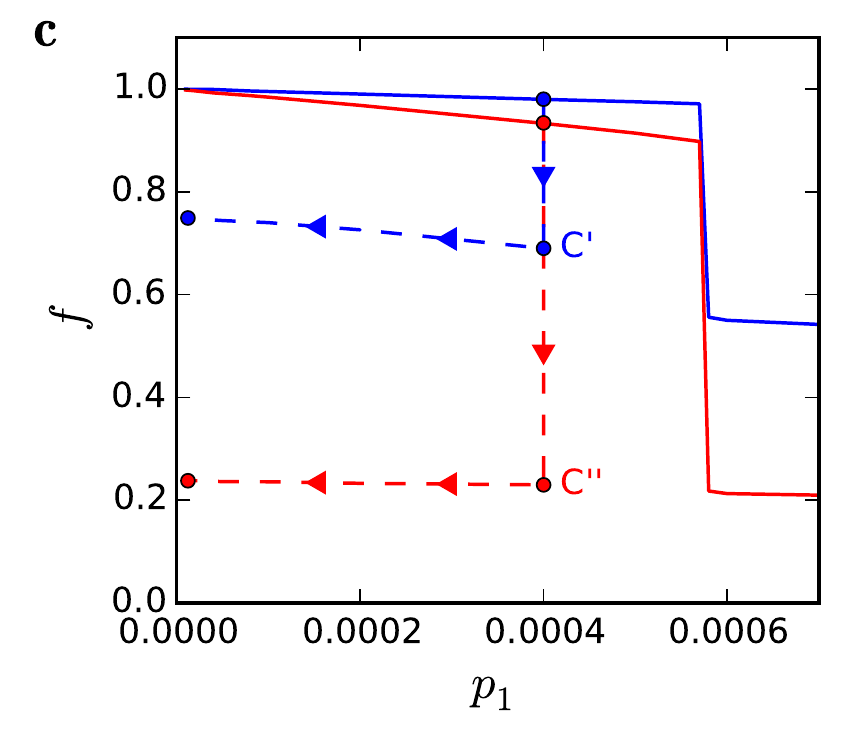}
\centering \includegraphics[width=0.23\textwidth]{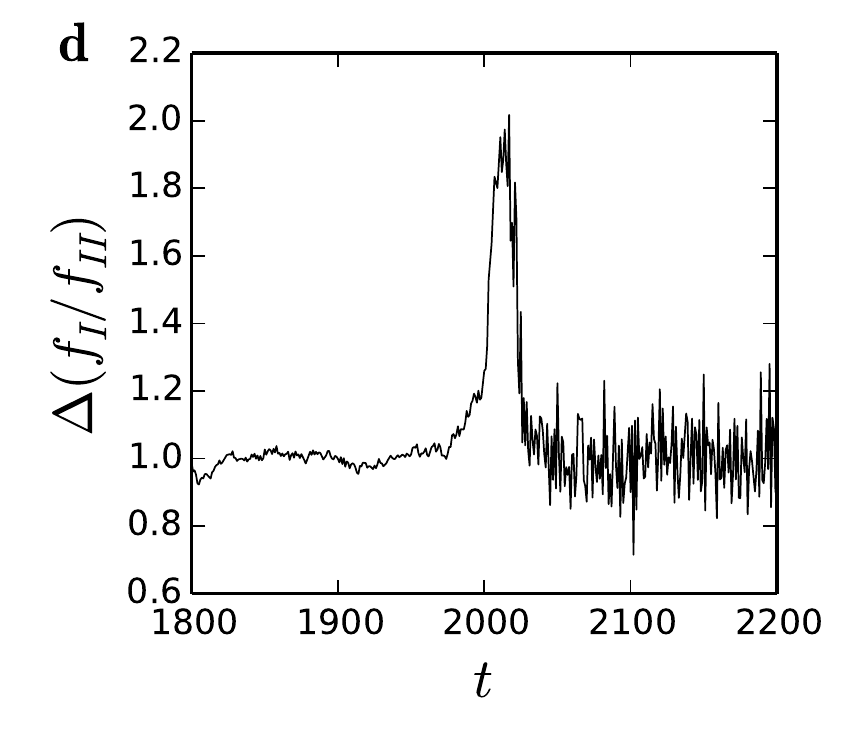}
\caption{ Quantification of the optimal attack duration.  Model
  parameters are: $m_S = 6, m_W = 3$, $m_{S,W} = m_{W,S} = 2$.  Similar as
  in Fig. \ref{33} but now with two thresholds used for defining
  resilience of each network.  External failure thresholds are $T_W =
  0.7$, $T_S = 0.3$, and $T_{S,W} = 0.4$.  In (a), we find the phase
  diagram in model parameters $(p_1,p_2)$ where each network has its own
  hysteresis. The takeover mechanism is disregarded and recovery time is
  $\tau=50$.  In (b), for each network we show the time evolution of
  the fractions of active nodes with the takeover mechanism included
  where we use a takeover period of $1.5 \tau$, $p_1=0.0004$, and
  $p_2=0.6$. At some point, S creates larger damage in a weak (less
  resilient) network W than to itself.  (c) Related with (b),
  as a result of the attack, both networks are damaged, but W is damaged
  more.  We also show how the fractions change with decreasing $p_1$
  during the recovery phase.  (d) Early-warning signal for
  determining when the attack should be stopped, defined as the change
  in the ratio between two fractions, $f_I(t+\Delta t)/f_{II}(t+\Delta
  t) - f_I(t)/f_{II}(t)$, where $\Delta t = 20$.  The attack should be
  stopped when the indicator reaches the maximum.  }
\label{3}
\end{figure}

Figure~\ref{3}(a) shows, for a given set of parameters, a two-parameter
phase diagram of competing networks, a model that incorporates  the
threshold separation for external failure but excludes takeover and feedback
mechanisms.  This model resembles that in Fig.~\ref{1} but utilizes
different configurations.  Suppose network S spontaneously activates at
time $t_0$ but, due to differences in the variables characterizing
network S and network W, initiates a substitution mechanism, not a
takeover.  Thus each time node $n_i$ in network W spends a time period in
an inactive mode that exceeds the substitution time
---e.g., in ecology,
a time period without food---
$n_i$ is replaced by a new node from network S.
Figure~\ref{3}(b) shows the fraction of active nodes in each network
calculated relative to the initial number of nodes at time $t_0$.
Fractions of active nodes of both networks exhibit a catastrophic
discontinuity (a phase flip) at $t \approx 2,000$, which is
characteristic of a first-order transition.  Since both networks are
interdependent, substituting nodes from the less resilient network W
affects the functionality of network S even more dramatically than that
shown in Fig.~\ref{1}.  Thus beyond some threshold we expect that
additional weakening of network W will also permanently damage network
S.  This demonstrates how dangerous an attacking strategy can be for an
attacker in a system of interdependent networks, e.g., between countries
 that are  at the same time competitors and economics partners.

Figure~\ref{3}(c) shows that when the attacks and substitutions cease,
the fractions of active nodes in network S and network W reach points
$C'$ and $C''$, respectively. If the probability of internal failure
$p_1$ spontaneously decreases during the recovery period, because of
network interdependence the functionality of network S is not
substantially  improved. The triumph of network S
over network W has its price.
In ecology, for example, although the
population of each species tends to increase, a dominance strategy is
risky, e.g., the extinction of a key species can trigger, through a
cascade mechanism \cite{Mold86,buldyrev_n10}, the extinction of many
other species \cite{Estes}.

Figure~\ref{3}(d) shows the change in the ratio between the fraction of
active nodes in network S and network W as a function of time. This
ratio can serve as an early-warning mechanism \cite{Dakos} that
indicates when attacks should be stopped.  Optimally, the stopping time
for attacks will be when the ratio reaches its maximum.

\begin{figure}
\centering \includegraphics[width=0.23\textwidth]{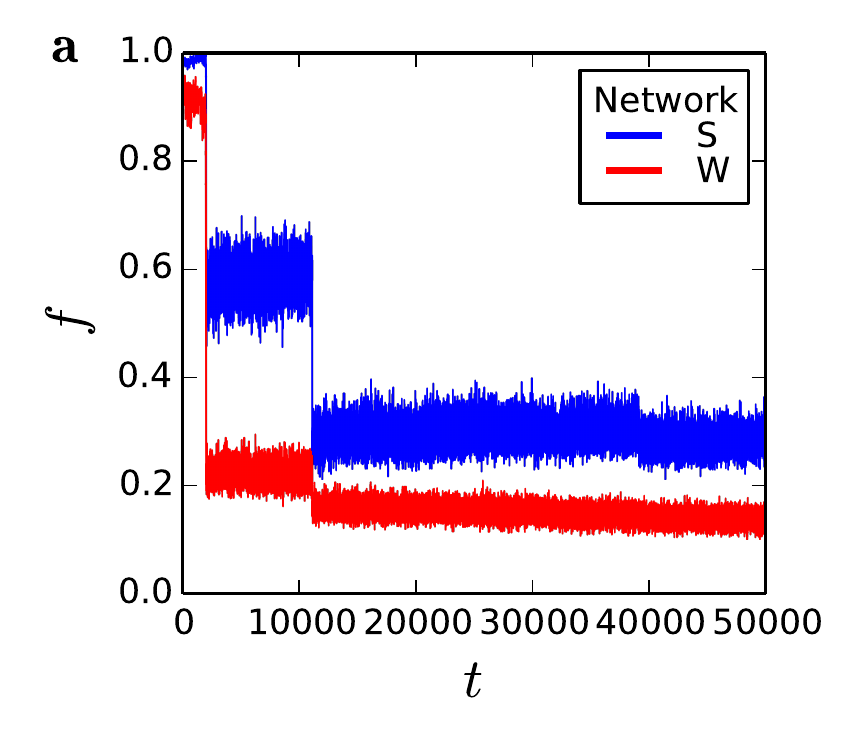}
\centering \includegraphics[width=0.23\textwidth]{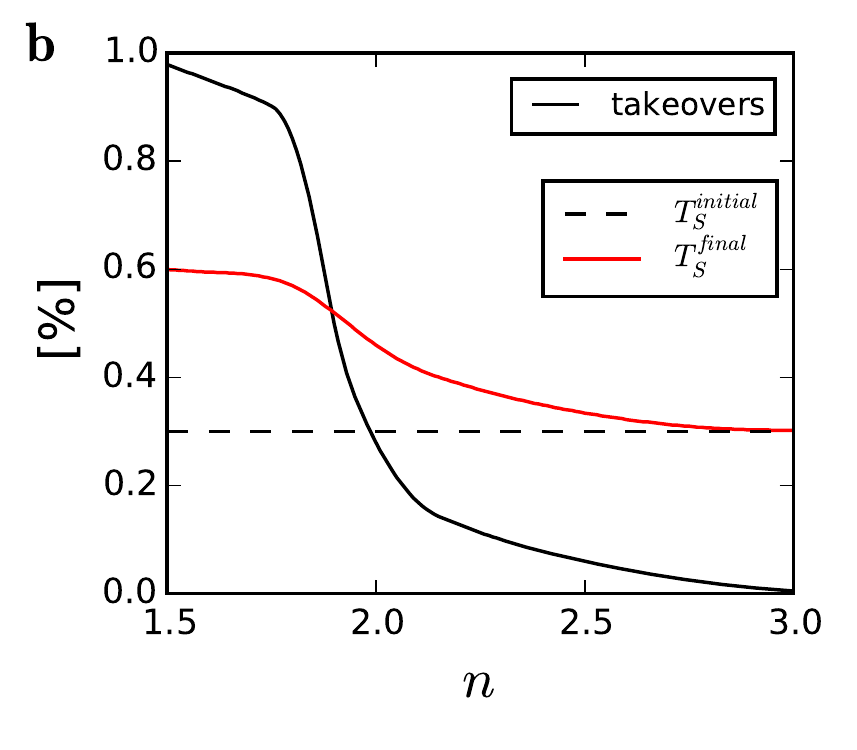}
\caption{ Evaluating the effect of the cost mechanism.
Model parameters
are: $m_S = m_W = 3$, $m_{S,W} = m_{W,S} = 2$, $p_1=0.0004$ $p_2=0.6$.
(a) The cost of the attacking strategy with takeover mechanism
additionally decreases network resilience. The fraction of active nodes
exhibits more discontinuities than in the case where the cost of an
attack was excluded (Fig. 5).  This is a consequence of the larger
change in the resilience of S due to inclusion of the cost mechanism.
(b) The fraction of takeovers and threshold $T'_S$ of the stronger
network S as a function of takeover time $n \tau$, with $\tau=50$.}
\label{4}
\end{figure}

Finally, Fig.~\ref{4}(a) shows that when the feedback mechanism (the
cost of taking over) defined in Eq.~(\ref{cost}) is included, the
fraction of active nodes in each network exhibits an even richer
discontinuous behavior than in Fig.~\ref{3}(c), where the cost was
excluded.  After 50,000 steps, because of the decrease in network S's
resilience after each substitute, the final fraction of active nodes in
network S is substantially smaller than the corresponding fraction in
Fig.~\ref{3}(c) (i.e., when the cost is excluded).  At the same time,
Fig.~\ref{4}(b) shows that an increase in the takeover time $n \tau$
decreases the fraction of substitutes.

\section{Summary}

In conclusion, we have presented a theoretical framework based on
resilience, competition, and phase transitions to introduce a
cost-of-attack concept that relates feedback mechanisms to resilience
dynamics defined using a linear conservation law.
Our model for
competing networks can be applied across a wide range of human
activities, from medicine and finance to international relations,
intelligence services, and military operations.

We focus on a specific context where one more resilient network attacks the
less resilient competitor network.
The model assumptions about the structure and dynamics for two interactive networks
with competing interactions and different resilience levels have to be adjusted in regard to
different real world scenarios (see the electronic supplementary material, S4).

The ability to measure
attacker network resilience and its attack cost is crucial because every weakening of
the resilience reduces the probability of the network survival under future
attacks.
For example, in political socio-economic systems a network-based approach for
overcoming competing countries could be more effective by applying economical sanctions than carrying out military actions.
Interdependent links established between countries
during prosperous times can facilitate sanctions (intentional
fluctuations) that are used as a weapon when more resilient countries
try to overcome less resilient countries.
They can also facilitate the
global propagation of economic recessions (spontaneous
fluctuations). During long economic crises these interdependent links
can become fatal for less resilient countries, whose weakness is enhanced by being underdogs in a global network-of-networks and, at the same time,
whose resources can be captured by more powerful countries.

Although, our proposed framework is suited for representing the most simplest case
of bilateral economic interdependence between just two countries (networks),
it provides the basis for more general scenarios
of alliances of more countries (networks). The
concept of alliance where some countries unite in order to attack some other
alliance is especially interesting when there is heterogeneity
in resilience of allied attacker countries. For example,
economically most dominant countries can increase their dominance at the expense
of their partners in the alliance or they can, on the other hand, depend on the alliance's weakest country
(see the supplementary material, S4 A).

In addition to the
intentional fluctuations characteristic of human societies, our
methodology can also be applied to a broad class of complex systems in
which spontaneous fluctuations occur, from brain functioning to
ecological habitats and climate fluctuations
\cite{Nowak_11,Adger,Cavoglave,Lange,Hunter,Estes,Vespignani12,Guido13}.
The methodology is based on specific structure, dynamics and mechanisms of the model of networks with competing interactions
and different resilience levels, that have to be adjusted for different systems and contexts of application (see the supplementary material, S4).

\clearpage

\noindent \textbf{Acknowledgments} \\
We thank Jacobo Aguirre and David Papo for discussions.

\noindent \\ \textbf{Author contributions} \\
B.P., D.H., T.L., M.P., J.M.B. and H.E.S. conceived and designed the research. B.P., D.H., T.L. carried out the numerical simulations, analysed the results and developed the theory. All authors discussed the results and contributed to the text of the manuscript.

\noindent \\ \textbf{Funding Statement} \\
B.P. work was partially supported by the University of Rijeka. M.P. acknowledges support from the Slovenian Research
Agency (Grant P5-0027), and from the Deanship of Scientific Research, King Abdulaziz University (Grant 76-130-35-HiCi).
J.M.B. acknowledges financial support of MINECO (project FIS2013-41057-P).
The Boston University work was supported by ONR Grant N00014-14-1-0738, DTRA Grant HDTRA1-14-1-0017 and NSF Grant CMMI 1125290. The authors declare no competing financial interests.

\clearpage

\setcounter{page}{1}
\renewcommand\thepage{\roman{page}}

\setcounter{section}{0}
\setcounter{figure}{0}
\renewcommand\thesection{S\arabic{section}}
\renewcommand\thefigure{S\arabic{figure}}
\renewcommand\thetable{S\arabic{table}}

\onecolumngrid
\begin{center}
\begin{large}
\textbf{Supplementary material: The cost of attack in competing networks}
\end{large}
\end{center}

\twocolumngrid

\section{Resilience dynamics in finance}

Here we demonstrate that the resilience dynamics in finance
  can be presented in
terms of a conservation (linear) law, where the threshold is controlled
by an asset-debt ratio.  Recall that in our model a node externally
  fails with a
certain probability when the total fraction of its active neighbors is
equal to or lower than the fractional threshold $T_h$.  The larger the
$T_h$ value, the less resilient the network.  In quantifying the impact
of a perturbation (attack) on the network, we shall demonstrate
  that the more severe
 the attack, the larger the impact on the network resilience.

 Suppose a bank $n_i$ has an interbank asset
$A^B_{i}$ invested equally in each of its $k_i$ neighboring banks.  Bank
$n_i$ has also some  asset $A_{i}$, considered as a stochastic
variable.  Following Refs.~\cite{Gai10,EPL12S}, we define a
bank to be solvent (active) when $(1 - \phi) A^B_{i} + A_i - L^B_{i} > 0
$, i.e., when the bank's assets exceed its liabilities, $L^B_{i}$.
  Here $\phi$
represents the fraction of inactive neighboring banks that
$n_i$ can withstand
and still function properly. Note that this $\phi$ is related to threshold $T_h$
as $T_h = 1 -\phi$,  since we assume that the number of incoming links is equal to the
number of outgoing links, $A^B_{i} = L^B_{i}$. The larger the
$A_i$ value, the more stable the bank.  Suppose that for each bank there
is a linear dependence between $A^B_{i}$, $A_i$, and the
network degree $k$---e.g., that $A^B_{i} = k_i$ and $A_i = 0.5 k_i $.
Then a bank is inactive when $1 - T_h = \phi = 0.5$ ($T_h = 0.5$) or
when at least 50\% of its neighboring banks are inactive. Let us assume
 next that $A_i$
increases due to an external perturbation $A_i = 0.5 + \epsilon$. Then
the new threshold is equal to
\begin{equation}
T'_h = 0.5 - \epsilon / k_i,
 \label{th}
\end{equation}
 or
\begin{equation}
    k_i ( T'_h -  T_h)  =  - \epsilon.
    \label{cost1}
\end{equation}

If the external perturbation is negative (positive), or alternatively,
 the asset increases (decreases) (Eq.~(\ref{th})),
the threshold
increases (decreases) and the resilience decreases (increases).  The
larger the number of neighbors ($k_i$), the smaller the change in the network resilience.  Note that if the external perturbation attacking one node
is shared not only by its neighbors but by the entire network with $N$
nodes, then $k_i$ is replaced by $N \langle k_i \rangle$ as in
Eq.~(4) in the paper. Note that if we replace the linear dependence between
assets and degree with a non-linear (power-law) dependence, e.g.,
$A^B_{i} = k_i^{\alpha}$, where $\alpha$ is a constant parameter, we obtain a
similar relationship to Eq.~(\ref{cost1}), $ k_i^{\alpha} ( T'_h - T_h)
= - \epsilon$.

\section{Resilience dynamics in interconnected networks}
We focus
on a mean field approximation of the level of external and internal
failures between nodes for two interconnected networks. Every node has an internal failure probability
 $p_1$~\cite{Antonio04S}---assumed, for reasons of simplicity, to be the same in both
networks, $p_{1,S}=p_{1,W}=p_{1}$.
If each node in network S has $k_S$ links with nodes in its
own network and $k_{W,S}$ links with nodes in network W, here we define that there must be
more than $m_S$ nodes in network S and $m_{W,S}$ nodes in network W if the
nodes in S are to function properly. In contrast,  if the number of inactive nodes in
network S is $\leq m_S$, the probability that the node in network S will
externally fail is $p_2$. Similarly, if the number of inactive
nodes in network W is $\leq m_{W,S}$, the probability that the node in
network S will externally fail is $p_2$. To simplify we can use
 $p_{2,S}=p_{2,W}=p_{2}$.
  We denote the time averaged
fractions of failed nodes in network S and network W as $a_S$ and $a_W$,
respectively. Using combinatorics, we calculate the probability that a
node in network S will have a critically damaged neighborhood among its neighbors in S to be $E^S
= \sum_{j=0}^{m_S} {k_S \choose k_S - j} a_S^{k_S - j} (1 -
a_S)^j$. Similarly, we calculate the probability that a node in network
S will have a critically damaged neighborhood among its neighbors in W to be $E^{W} =
\sum_{j=0}^{m_{W,S}}{k_{W,S} \choose k_{W,S} - j} a_W^{k_{W,S} - j} (1 -
a_W)^j$. The probability that a node will fail externally due to
failures in network S (network W) is $p_2 E^S$ ($p_2 E^W$).  If we
denote the internal failures in network S by A, and the external failures
by B, and the external failures in network W by C, then the probability
that a randomly chosen node in S will fail is $a_1 \approx P(A) + P(B) + P(C)
- P(A) ( P(B) + P(C)) - P(B) P(C) + P(A) P(B) P(C)$. If we assume that
A, B, and C are not mutually exclusive, but interdependent
events, we come to
\begin{eqnarray}
a_S &\approx &p^*_1 + p_2 (E^{S} + E^{W}) -  p^*_1 p_2 (E^{S} + E^{W}) \\
&-&
p_2 p_2 E^{S}  E^{W} + p^*_1  p_2  p_2 E^{S}  E^{W}.
\end{eqnarray}
 where $p^*_1 = 1 - \exp(-p_1 \tau)$,  and  as in Ref.~\cite{Antonio04S}
  node $j$ recovers from an internal failure after a time
period $\tau$.
Similarly, from the above equation we obtain the fraction of failed nodes $a_W$ in network W (either internally or externally failed) by interchanging S and W.

\section{Coupled BA interdependent networks with equal connectivity
    but different threshold}

\begin{figure}[b]
\centering \includegraphics[width=0.24\textwidth]{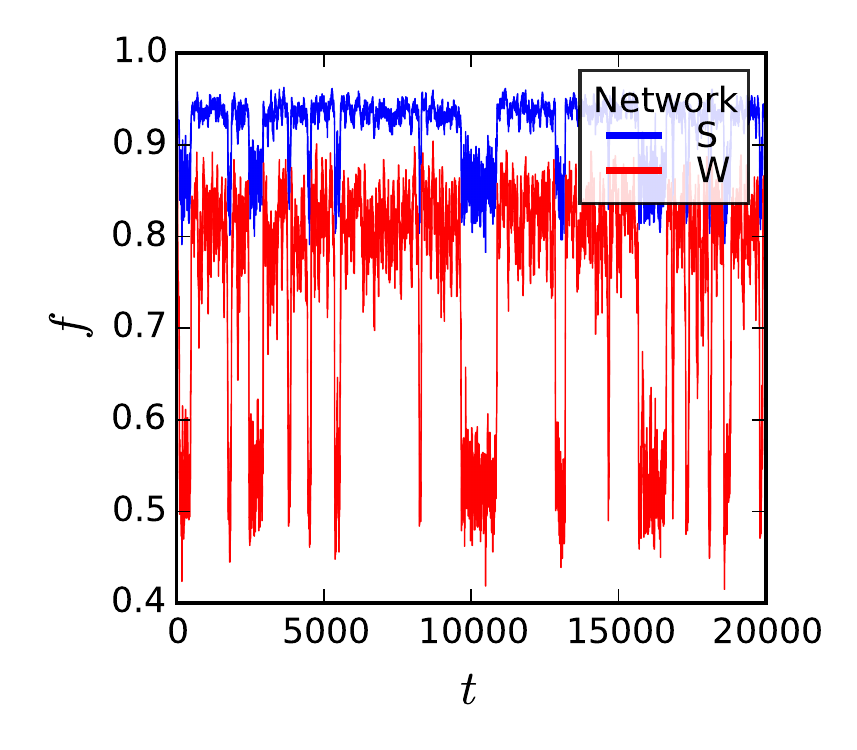}
\caption{ Coupled interdependent networks with equal connectivity
    but different threshold.  For the parameters we use: $\tau = 50$,
  $p_1=0.0012$, $p_2=0.48$, $m_{S}=m_{W}=3$,  $m_{S,W}=m_{W,S}=2$, $T_S =
  0.3$, $T_W = 0.7$.  The larger the
$T_h$ value, the less resilient the network.
 We obtain that both networks exhibit the same
  hysteresis. The parameters are set in a part of phase space
   to enable phase flipping between
  active and inactive states.  However, due to different thresholds,
   network S is more resilient than  network W---the average fraction of
  active nodes in  network S is practically always larger than in  network
  W.  }
\label{7}
\end{figure}

We now focus on two interdependent BA networks with equal connectivity
    but different threshold, where $T_S < T_W$.
Fig.~\ref{7} shows that for each of the
 two interconnected networks the
fraction of active nodes  simultaneously jumps from
 a stable state to another one. Phase-flipping is  obtained by setting
  the network close to a critical point that is reported for a single
   network in Ref.~\cite{Antonio04S}.
Since the threshold in network S is substantially smaller than in
network W, and so S is more resilient than network W, the fraction of
active nodes in network S is larger than  the fraction of active nodes
in network W. Thus the volatile
 phase-flipping in network functionality is more dangerous for network W
  than for network S.
The fractions, as a function of time, can approximately model the
populations of preys and predators and so be related to the periodic
solutions of the Lotka-Volterra
(predator-prey) equations \cite{Nowak_11S}.

\section{Applicability of the model to real world scenarios}

Most real world systems are composed of networks which interact with each other in different ways.
In our model we focus on a specific context where one more resilient (strong) network (attack initiator) attacks the less resilient (weaker) competitor network (attack target).
The proposed model depends on the structure of two interactive competing networks and
the defined dynamics between initiating an attack against a competitor network and the consequences that might weaken the attack initiator resilience.
Within this framework assumptions about the structure and dynamics for two interactive networks with competing interactions and different resilience levels have to be adjusted in regard to
different real world scenarios.

For the structure of interactive networks we use multilayer network model distinguishing between intra-network and inter-network connections. Although in our analysis we focused on specific architectures of intra-network and inter-network connections, in general they can be arbitrarily chosen depending on real world examples. Each competing network is formed by the intra-network connections between its own nodes, while interactions between competing networks are determined by inter-network connections between nodes from two opposing competitor networks. We also assumed static intra-network and inter-network connections, while in different real world examples these connections can also change through time and depend on different spatial and temporal attributes.

When we represent both an attacker or attacked system as a network, we must also decide which attributes and level of details of the system are of interest for a specific real-world scenario. A single node in network represents a particular component of a system or aggregated set of similar components, while the consequences of an attack are described by an active or an inactive node's state.
The global collective state of both the attacker and the attacked network is measured by the total fraction of active nodes in each network. Finally, the interpretation of node's activity and inactivity depends on the application context of different real world attack scenarios.

In order to explain the general network vulnerability to global cascades of local and external node failures/inactivations caused by attack consequences, we model node dynamics by a cascade contagion model with stochastic internal and external activation/inactivation process \cite{Antonio04S}, inspired by the Watts threshold model \cite{Watts02S} where each node has a probability to internally fail/inactivate independently of other nodes. At the same time, the probability to externally fail/inactivate if a fraction of its active intra-network and inter-network neighbors is less than or equal to a specified fractional threshold. This fractional threshold can be interpreted as the network resilience to failures due to interdependencies, the larger the value of the fractional threshold, the less resilient the network.

We assume that the attacker network has a higher resilience than an opponent network and that it is willing to partially weaken its own resilience in order to more severely damage a less resilient competitor. Attack concept and also its influence on resilience dynamics of the attacker is based on causing critical fraction of active neighbors for nodes from an opponent network, and using takeover and substitution mechanisms for them after their long inactivity. However, this concept is highly context-dependent and its mechanisms have to be properly mapped in regard to different real work scenarios.

As the attacker network has higher resilience than an opponent network it can induce the implicit indirect attack by increasing the probability of internal failures/inactivations of its nodes causing critical fraction of active neighbors for nodes from an opponent network. In addition, takeover and substitution mechanisms define direct attack on an opponent network. Takeover mechanism is related to the explicit attack on an opponent network with the same type of nodes as the attacker network and the explicit reduction of the attacker network resilience by increasing its fractional threshold. On the other hand, the substitution mechanism represents the explicit attack on an opponent network with implicit reduction of the attacker network resilience by having nodes whose state is more dependent on nodes form the opponent networks.

Although we focus on the most simplified scenario of two interactive competing networks where just one and more resilient network can attack a less resilient opponent network, the introduced framework provides the basis for more general scenarios
of alliances of more different networks with also different interaction types between them (e.g. competing intra-specific/network, antagonistic, mutualistic) and strategies for attack as well as defense. How many new parameters have to be introduced for this scenario also depends highly on the context of the study of interest. We can observe each network collective state and resilience dynamics separately or depending on their affiliation while a specific attack or defense strategy is taking place.

How successful our network model might be when applied in practice depends first on how capable we are in estimating model parameters and mapping them to real-world context. As described before our model is structured with the architecture of intra-network and inter-network connections which define networks structure and constraints for an interaction, and with parameters for cascade dynamics including internal and external failure probabilities, the fractional threshold and the limit for inactivity period. We show how these parameters can be mapped to two real world scenarios from ecology and political socio-economy application domains.


\subsection{Socio-economic system: competing inter-firm network of countries under economic sanctions }

Here we demonstrate how our modeling framework is mapped to a specific political-socio-economic scenario where one country imposes economic sanctions or
similar political-economic attack/action against another country. In this regard we explain interpretation of model parameters, intra/inter-network architecture and activation/inactivation dynamics.
A critical aspect of political-economic attack/action is the economic costs endured not only by the targeted country, but also by the sender country (the one making the action).
Although, we are focusing here on the most simplest case of bilateral economic interdependence between just two countries, such actions
can also bring to economic consequences to other third parties like neighbours countries of the target or its trading partners, but as well as senders.
Economic sanctions are deliberate, government-inspired withdrawal, or threat of withdrawal, of customary trade or financial relations and
are most effective when aimed against friends and close trading partners as these countries have more to lose than countries with which the sender has limited or adversarial relations.

Interdependent links established between countries
during prosperous times can facilitate sanctions (intentional
fluctuations) that are used as a weapon when more resilient countries
try to overcome less resilient countries. They can also facilitate the
global propagation of economic recessions (spontaneous
fluctuations). During long economic crises these interdependent links
can become fatal for less resilient countries.
Similarly, when economic sanctions are lifted for the weaker economies is much harder to restore than
for the stronger economics which recover after suffering little damage.

In order to explain the economic cost for the country imposing the sanctions,
we first specify state of the country's economic system by collective activity of individual firms which are part of the country economy.
Firms are connected to each other directly or indirectly through their business transactions
(i.e. obtaining materials from suppliers, delivering goods/products to services or R\&D cooperation).
Economic network of each country is represented by inter-firm network assuming
that intra-network connections specify cooperative interactions between firms of same country, and
inter-network connections specify interactions between firms in different countries that are subjected to
the bilateral economic relation between two countries.

As many firms borrow from and lend to each other, and in particular
when these firms are speculative and dependent on the credit flow,
shocks to the liquidity of some firms may cause the other firms to also experience financial difficulties \cite{Pod14S}.
The way how geographically localized shock propagate through such inter-firms networks
determines country resilience to the shock experienced by their corresponding firms.
These dynamic vulnerabilities of firms can be related to the fractional threshold parameter that can be controlled
by its asset-debt ratio where activity/inactivity of a firm is determined by its solvency (see S1. Resilience dynamics in finance).

The attack concepts is related to inducing the implicit indirect attack by increasing probability of internal failures/inactivations of its firms causing target country firms to become insolvent.
In addition, takeover mechanisms define attack on the target country economy by taking over their firms which are insolvent for some critical time.

The extension of our modeling framework from bilateral economic interdependence between just two inter-firm networks of competing countries to more general scenarios of network alliances
where some countries unite in order to attack some other alliance is especially interesting when we introduce the resilience heterogeneity of allied attacker countries.
The resilience heterogeneity of allied attackers can induce opposing interests and attack strategies between partners as well as provide descriptions of scenarios where
economically most dominant countries increase their dominance relative to other countries and at the expense of their partner countries in the alliance or where, on the other hand,
economically strong countries depend on alliance's weakest country. For example, suppose two countries, I and II  imposing economic sanctions at the third country, III,
where I is more resilient than II and both I and II are more resilient than III. A country resilience can be different since II is more linked with III than I.
When the sanctions are lifted, the relative resilience between I and II can be increased making I even more dominant
than before the sanctions. But on the other hand due to the interdependencies between allied countries, i.e. countries I and II, attack strategy of I is dependent also on the state of II.

\subsection{Ecological system: competing animal social networks under predator-prey interactions }

Here we demonstrate how our modeling framework is mapped to two competing animal social networks under predator-prey interactions. We explain possible choices and an interpretation for intra/inter-network structures and activation/inactivation dynamics for a specific scenario where individuals from one predator species attack individuals from one prey species. A predator is an animal that hunts and kills other animals (its preys) for food. An interpretation of predators and preys internal or external activity depends on the given context of the study of interest and the interpretation of their intra/inter-network connections.

The intra-network structure for attacking (predator) and attacked (prey) networks can be represented by corresponding socio-spatial network of their individuals \cite{Pinter-Wollman13,Farine15,Wilson15,Kurvers2014}. While, inter-network connections, i.e. connections between predator and prey individuals, determine which predator individuals can come in close contact with which prey individuals to hunt them based on spatial and temporal factors. Interactions between animal individuals are highly dependent on their spatial proximity. Thus determining the structure of social animal network also depends on given spatial context \cite{Pinter-Wollman13}. Recent technological innovations in tracking devices and reality mining approaches are starting to enable remote monitoring and collection of detailed information of behaviors of individual animals at high spatial and temporal resolution \cite{Wilmers2015,Krause2013,Handegard12}.

A structure of social network between animal individuals influence diversity of social behaviors such as finding and choosing a sexual partner, making movement decisions, engaging in foregoing or anti-predator behavior which is manifested at the population level in the form of habitat use, mating systems, information or disease transmission. Social interactions between individuals of same species can differ in their type (competitive, cooperative, sexual), frequency and duration. Animals modify their social interactions in response to changes in external conditions such as climate, predation pressure, and social environment \cite{Pinter-Wollman13}. Although, interactions among animals are dynamic, many animal social network studies examine static structures.

For a simplified predator-prey interaction scenario we can assume static cooperative interactions and that there are no competing interactions between individuals of the same species, i.e. infraspecific competition is not taken into account. Cooperating interactions between individuals play an important role in the spreading of information within the network affecting an access to resources and the probability of predation of an individual. For example, individuals that have large number of intra-network connections can be better in discovering new food patches.

With that interpretation of the intra-network and inter-network connections, we can model predators coordinated attack by their activation dynamics. In this context it is reasonable to assume that a node in attacked (prey) network will externally fail (be killed by predators) if there is enough activated predators (i.e. fraction of active inter-neighbors larger then given threshold) that are in close range of the prey (i.e. those connected with inter-network connections). As usual, prey can also externally fail (die or migrate) if its fraction of active intra-neighbors is equal or smaller then the fractional threshold.
In this context we can interpret internal and external failure/activation probabilities, and the limit for inactivity period for predators as their opportunity, exigency and willingness to hunt prey together in certain time. While those parameters in prey's case can be interpret as a way how prey sustains predators attacks or keeps a certain habitat patch active with its own species through reproduction. Prey organisms that are difficult to find, catch, kill or consume will survive and reproduce.

The explained attack concept is in contrast to the indirect attack concept where inactive/failed states of attacker nodes cause external failure of attacked network nodes. This assumption is reasonable in a scenario where inter-network connections represent prey dependencies on its predators, i.e. where for each prey there is a needed fraction of active predators for them to survive because the active predators balance the population size of a prey (see also section S3. Coupled BA interdependent networks with equal connectivity but different threshold). In that scenario, predators and prey depend on each other in that the predators rely on prey as food source, but in turn they also keep prey population against over-population which could cause decrease of their food source. Beside that, in some cases prey species can even facilitate its own predator \cite{Aguera15}. A possible context that relies on predator and prey interdependence is explanation of how a disease or other internal disturbances in predator population can effect prey population dynamics and vice versa.

In contrast to socio-economic scenario of competing network of countries' firms, takeover mechanism is not suitable attack concept for scenario of predator-prey interactions where we model interaction between social network of different node types, but substitution mechanism is suitable to represent additional level of predation influence on prey population. In general predation influences organisms at two ecological levels. At the level of the individual, the prey organism has an abrupt decline of its lifetime reproductive success, because it will never reproduce again, and at the level of the community, predation reduces the number of individuals in the prey population. Accordingly, the direct cost for strong predator which easily hunts and kills its prey is generally negligible, but indirect cost can be manifested through the fact that each reduction in prey abundance impedes finding a next prey. On the other hand, the direct cost is noticeable for intraguild predators, predators that kill and eat other predators of different species at the same trophic level \cite{Barraquand13,Borrelli15}.

Classical models of predator-prey population dynamics, often considered as the basic building blocks of larger, food-web models, explain how tropic interactions lead to oscillatory population cycles. These models assume well-mixed population where all pairs of individuals have equal probabilities of interacting with each other which corresponds to a complete graph structure. This means that the encounter rate between predators and prey is expressed in a mass-action fashion, i.e. as a product of prey and predator landscape densities \cite{Barraquand13}. Our model can describe mechanistic approach of predator-prey interactions within structured population of individuals of two animal species.

Animal social networks are harder to determine and observe than human social networks as animal network data must be collected by direct observation of interactions between individual animals \cite{Wilmers2015,Krause2013}. For some large herding species it may be possible to count every individual, but for many species this is not possible.
In cases when is too time consuming to collect animal network data at individual level due to difficulties of capturing and identifying individuals, it can be useful to observe categories of individuals and consider interactions between them. Since spatial proximity is significant factor for establishing and maintaining cooperation between individuals, we can observe interactions between predators and prey through patches they occupy \cite{Kondoh15,Macdonald15}.

While in social animal networks, one node represents just a single individual of a given species  and each edge represents some form of interaction between two individuals, in food webs and ecological network studies a node typically represents a species while connections between nodes represent different types of interactions between species. Species may interact with each other through antagonism (prey-predator, host-parasitoid, or host-parasite interaction), competition, or mutualism \cite{Ings09}. Food webs can provide initial blueprints of inter-connections between animal socio-spatial networks when extending our modeling framework from interaction between just two species networks to more general scenario of network alliances where several predator and/or prey networks can cooperate or compete between each other. For example, in the case when two prey species have a common predator, one prey species can lead to indirect exclusion of the other species or the case where two predators alone compete for a single prey species, one species is always excluded by the other, even in the presence of a top predator.

\end{document}